\documentclass{article}
\usepackage{ltwol2e}
\usepackage{epsfig}

\newcommand{\pp}{{_{I\hspace{-0.2em}P}}}

\def\be{\begin{equation}}
\def\ee{\end{equation}}
\def\bea{\begin{eqnarray}}
\def\eea{\end{eqnarray}}

\begin{document}

\title{FRACTURE FUNCTIONS FOR DIFFRACTIVE AND LEADING PROTON DIS}

\author{D. DE FLORIAN}

\address{Theoretical Physics Division, CERN, CH 1211 Geneva 23, Switzerland
\\E-mail: Daniel.de.Florian@cern.ch}   

\author{R. SASSOT}

\address{Departamento de F\'{\i}sica, 
Universidad de Buenos Aires, 
C. Universitaria, Pab.1 
Buenos Aires (1428), Argentina
\\E-mail: sassot@df.uba.ar}

\twocolumn[\maketitle\abstracts{ We present a combined QCD analysis of diffractive and leading proton deep inelastic scattering data using the framework of fracture functions. It is shown that this framework allows a precise and unified perturbative QCD description
for the data, alternative to those that relay on Regge factorization.}]

\section{Introduction}

In recent years considerable attention has been paid to diffractive and leading-proton deep inelastic scattering processes \cite{ZLP,H1LP,H1D,ZD}. These two kinds of processes differ considerably in the kinematical regions where they are produced and also in the non-perturbative mechanisms or models used to explain them and, consequently, their description is usually done within completely unrelated frameworks and in terms of different structure functions: the 
leading-proton and the diffractive deep inelastic scattering structure functions, $F_2^{D(3)}$ and $F_2^{LP(3)}$, respectively.

However, if the experiment under consideration is able to identify the final-state proton in the diffractive processes, as in the ZEUS measurements, or  if there is confidence in the dominance of the single dissociative process $\gamma^*p \rightarrow Xp$, both kinds of events can be thought as of semi-inclusive nature, with identical final-state particles produced in the target fragmentation region. From this point of view, the perturbative QCD framework for their description must be identical, with only one factorized observable, even though the specific models for their non-perturbative features are completely different.

In perturbative QCD, the most appropriate description, for semi-inclusive DIS events in which the identified final-state hadron is produced in the target fragmentation region, is the one that includes fracture functions \cite{VEN}. These functions can  be  understood essentially as parton densities in an already fragmented target, and extend the more familiar description of semi-inclusive events in terms of parton densities and fragmentation functions,
allowing a leading-order description for target fragmentation events in the forward direction, and solving some problems asociated to factorization \cite{API,GRAU}.  At the same time, the formalism provides
not only the motivation for establishing a connection between the kinematical regions of these kinds of processes, but also rigorous predictions about the perturbative QCD behaviour of the cross-sections. A priori this behaviour  is not the same as that of the ordinary structure functions due to appearence of an inhomogeneous term in the Altarelli--Parisi equations for fracture functions \cite{VEN,CGT}.

In the following we present results from a QCD global analysis of  recent semi-inclusive deep inelastic scattering data produced by H1  using the framework of fracture functions. It is shown that this approach unifies the description of   diffractive and leading-proton phenomena and allows a perturbative QCD description without the usual assumptions about approximate Regge factorization. 
The resulting parametrization is also used to compute other observables measured by ZEUS, not included in the fit to H1 data, finding also an outstanding agreement with the data.

\section{Definitions}

It is customary to define the leading-proton structure function $F_2^{LP (3)}$
from the corresponding triple-differential deep inelastic scattering cross section:
{\small
\begin{equation}
\frac{d^3 \sigma^{LP}}{dx\,dQ^2\,d\xi} \equiv \frac{4\pi \alpha^2}{x\,Q^4}\left( 1-y+\frac{y^2}{2}\right) F_2^{LP (3)}(x,Q^2,\xi) \, ,
\end{equation}} 
and the usual kinematical variables.

Even though the processes accounted for in Eq. (1) are explicitly of a  semi-inclusive nature, the formulation based on the leading-proton structure
function is used instead of the usual approach for semi-inclusive deep inelastic  scattering in terms of parton distributions and fragmentation functions, because the last one only takes into account hadrons produced in the current fragmentation region and thus not contributing to the forward leading hadron observables. 

These problems, and those related to factorization of collinear singularities at higher orders, are overcome, however, if the complete perturbative framework
for semi-inclusive processes is taken into account, for which the LO expression for the production of very forward hadrons is
{\small              
\begin{eqnarray}
\frac{d^3\sigma^p_{target}}{dx\,dQ^2\,dz} =\frac{4\pi \alpha^2}{x\,Q^4}\left( 1-y+\frac{y^2}{2}\right)   \sum_i e^2_i x M_i^{p/p}(x,z,Q^2)
\end{eqnarray}}
where $M_i^{p/p}(x,z,Q^2)$ is the fracture function that accounts for target fragmentation processes and we define the variable $z\equiv E_{P'}/E_{P}$ as the  ratio between the energies of the final-state proton and the proton beam in the centre of mass of the virtual photon--proton system. For very forward protons then, $1-\xi \simeq z $.

Fracture functions can be thought of in terms of the elements of any lowest-order picture for hadronization, for example as the product of a flux of exchanged `reggeons' times their structure functions, or more formally as an ingredient of the perturbative QCD treatment: the non-perturbative parton distributions of a proton fragmented into a proton. The latter choice can be generalized to higher orders, allowing a consistent analysis of the scale dependence of the semi-inclusive  cross section. This scale dependence is driven by inhomogeneous Altarelli--Parisi equations, reflecting the fact that the evolution is not only driven by the emission of collinear partons from those found in the target (homogeneous evolution), but also by
the fragmentation of partons radiated from the one struck by the virtual probe (inhomogeneous evolution).

Defining the equivalent to $F_2$ for fracture functions, i.e.
\begin{equation}
M_2^{p/p}(x,z,Q^2)\equiv x \sum_i e_i^2 M_i^{p/p}(x,z,Q^2),
\end{equation}
and taking into account the shift from $z$ to $\xi$, the relation between this function and the leading-proton structure function is quite apparent.
In reference \cite{NEUTR}, the use of fracture functions for the description of leading-hadrons produced in the target fragmentation region has already been discussed in relation to the analysis of very forward neutrons observed by the ZEUS collaboration. There it was shown that, neglecting contributions beyond LO coming from the current fragmentation region, what is usually defined as the leading-neutron structure function $F_2^{LN (3)}$ is just the fracture function contribution to the semi-inclusive cross section. 

 Analogously to Eq. (1), the differential cross section for diffractive deep inelastic scattering is usually written in terms of the diffractive structure function $F_2^{D(3)}$ 
{\small        
\begin{equation}
\frac{d^3 \sigma^{D}}{d\beta\,dQ^2\,dx_{\pp}} \equiv \frac{4\pi \alpha^2}{\beta\,Q^4}\left( 1-y+\frac{y^2}{2}\right) F_2^{D (3)}(\beta,Q^2,x_{\pp}) 
\end{equation}} 
where $x_{\pp}\equiv \xi$, and the variable $\beta$ is used instead of $x$.
 When this last cross section is dominated by the single dissociative process $\gamma^*p \rightarrow Xp$, implying that there is a proton in the final state, the contributions to it are again given by the same fracture function in (2), even though in a completely different kinematical region.

Different kinematical 
regions correspond to different behaviours and also to different underlying 
models. The leading-proton structure function has been measured by the H1 collaboration \cite{H1LP}, and has also been compared with predictions of different  mechanisms, such as meson exchange and soft colour interaction models, implemented in event generator programmes, none of which reproduced integrally the main features of the data \cite{H1LP}. The standard interpretation for the diffractive cross section is given in terms
of `pomeron' exchange. In this framework, different model estimates, and even QCD-inspired parametrizations of the `pomeron' content,  have been proposed \cite{STIR,COLL} however the comparison between model predictions and data again  has been found to be rather poor unless a large number of additional elements is included \cite{H1D}.

In the language of fracture functions, both the leading-proton and the diffractive regimes are complementary features of a more general semi-inclusive process. The approach, then, suggests and provides a bridge between the two regimes, which is particularly appropriate and even necessary, at least in the kinematical region where neither the `pomeron' nor the `reggeon' exchange picture alone are expected to describe the data, say $x_{\pp} \sim 0.1$.

\setlength{\unitlength}{0.6mm}
\begin{figure*}[hbt]
\begin{picture}(340,155)(0,0)
\put(15,-10){\mbox{\epsfxsize7.7cm\epsffile{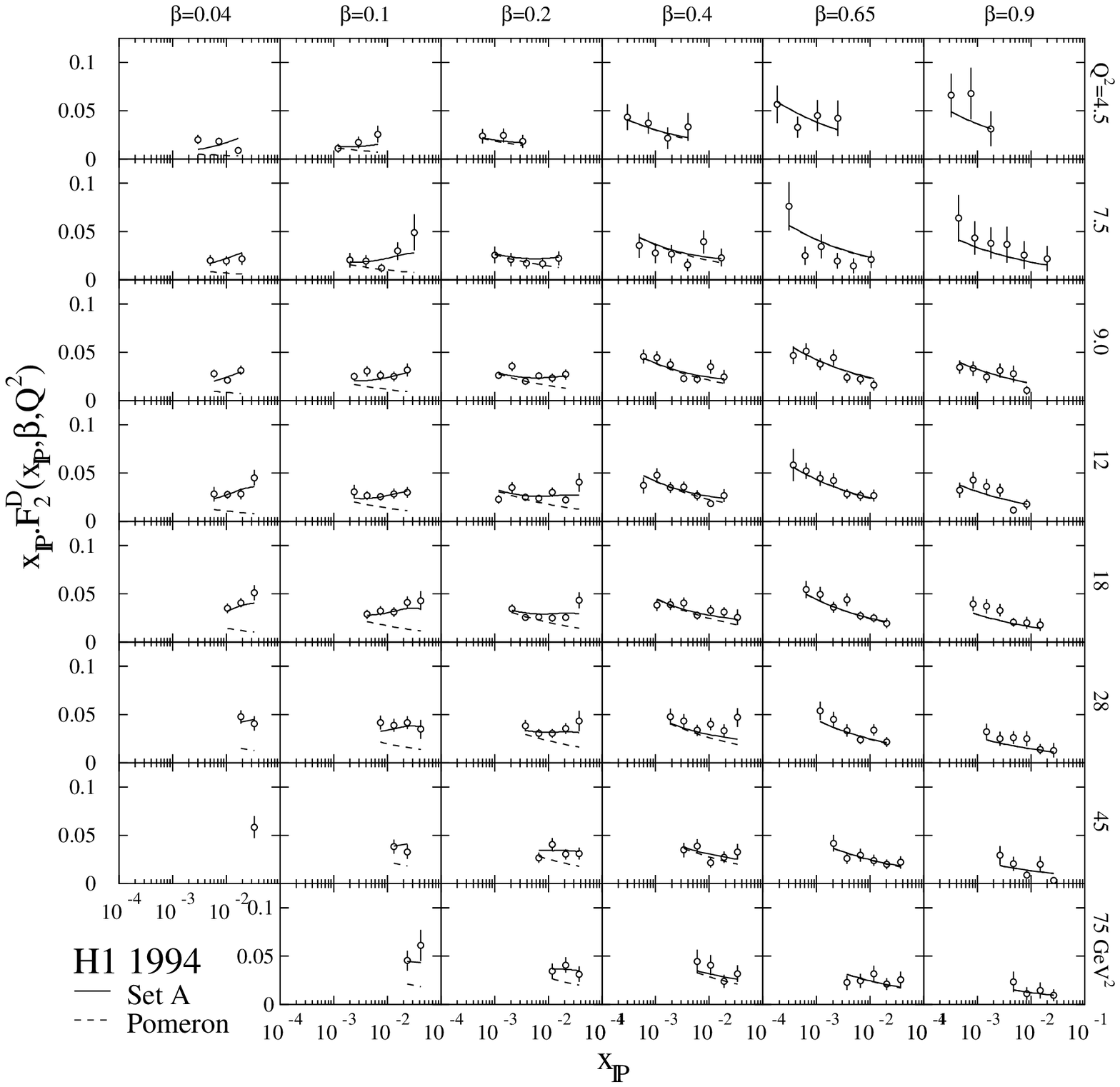}}}
\put(0,6){\mbox{\footnotesize Figure 1:  H1 diffractive data against the outcome of the fracture }}
\put(0,0){\mbox{\footnotesize function parametrization. }}
\put(175,-10){\mbox{\epsfxsize7.7cm\epsffile{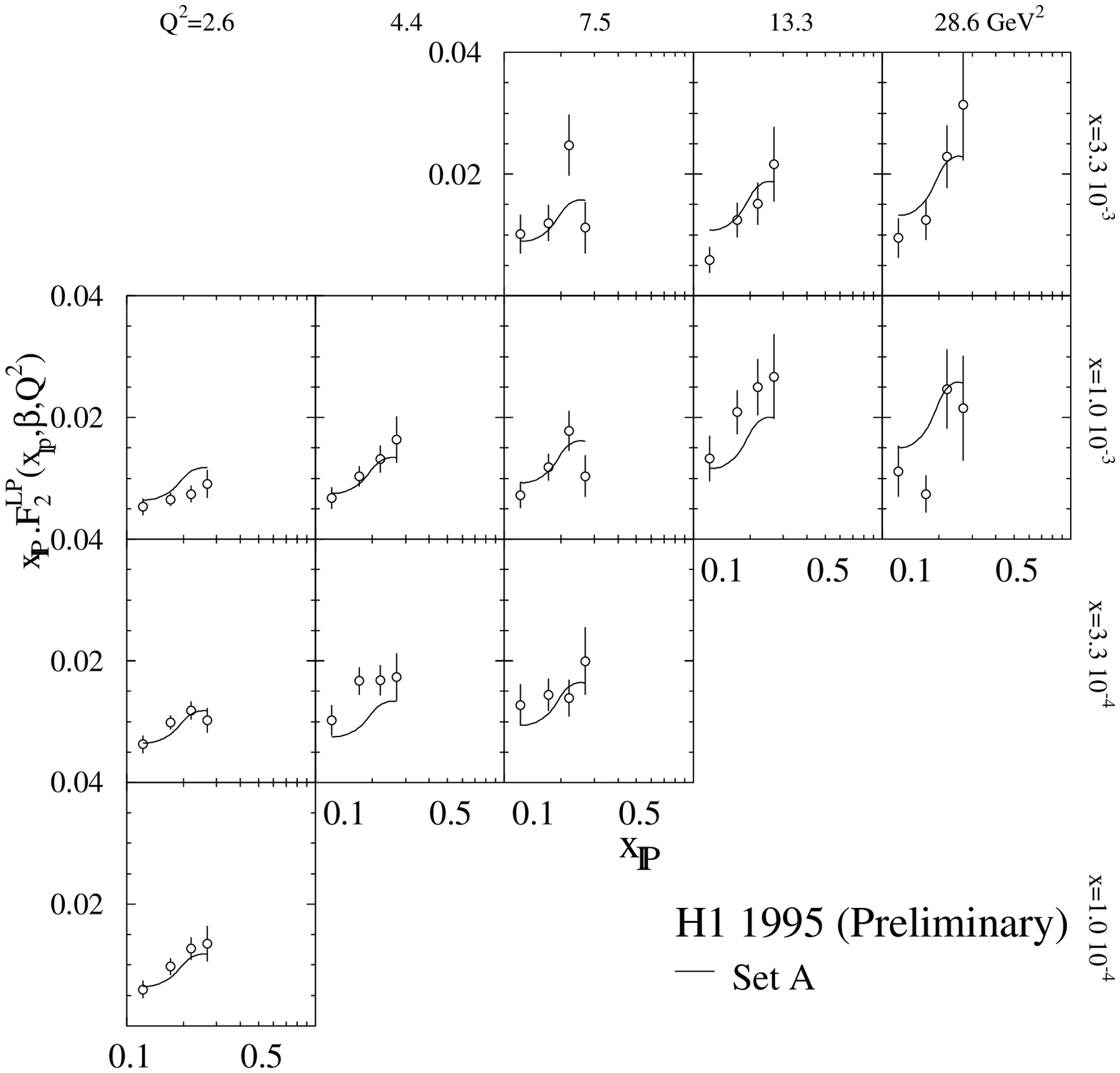}}}
\put(165,6){\mbox{\footnotesize Figure 2:  
H1 leading-proton data against the outcome of the }}
\put(165,0){\mbox{\footnotesize  fracture function parametrization. }}
\end{picture}
\end{figure*}

\section{Parametrization}

\begin{table*}[t]
\caption{Parameters for $Q_0^2=2.5\, $GeV$^2$. \label{tab:exp}}
\vspace{0.2cm}
\begin{center}
\begin{tabular}{|c|c|c|c|c|c|c|c|c|} \hline Set&$\alpha_{\pp}$&$C_{LP}$&$\beta_{LP}$&$\gamma_{LP}$&$a_{LP}$&$N_s$ &$N_g$ &$a_g$ \\ \hline
A & $-$1.260 & 14.395 & 32.901&2.627&12.320& 0.041& 0.354 & 0.450 \\  \hline
B & $-$1.257 & 12.556 & 32.412&2.338&11.412& 0.047& 0.694 & 0.648 \\  \hline \end{tabular} 
\end{center} 
\end{table*}

In order to obtain a parametrization for the proton-to-proton fracture function
$M_2^{p/p}(\beta,Q^2_0,x_{\pp})$ at a given initial scale $Q^2_0$, we  select in the first place, a relatively simple functional dependence in the variables $\beta$ and $x_{\pp}$. If one were only interested in the diffractive regime,
the natural choice would be a simple `pomeronic' flux in $x_{\pp}$ times an ordinary parton distribution in $\beta$. For the leading-proton regime the
natural choice would be almost the same, but with a standard meson or `reggeon' flux or even better, something combining their effects. These kinds of parametrizations give relatively good initial approximations to the description of the corresponding data sets; however their survival  seems unlikely  in a more precise analysis.

In order to take into account small departures from the initial approximations, and also combine the two behaviours in such a way that for  low $x_{\pp}$ (diffractive regime)  the `pomeron' picture dominates, while for low  $\beta$ and large $x_{\pp}$ the meson or `reggeon' exchange picture emerges, we propose a modified flux such that the light-quark singlet component $(M_q^{p/p} \equiv 3 M_u^{p/p}= 3 M_d^{p/p}=3 M_s^{p/p})$ of the fracture 
function is parametrized as
\begin{eqnarray}
x M_q^{p/p}(\beta,Q^2_0,x_{\pp}) =   \,N_{s}\, \beta^{a_s}(1-\beta)^{b_s}  \times \,\,\,\,\,\,\,\,\,\,\,\,\,\,\,\,\,\,\,\,\,\,\,\,\,\,\,\,\,\,\,\,\,\,\,\,\,\,\,\,   \\
 \left\{ C_{\pp}\, \beta \, x_{\pp}^{\alpha_{\pp}} + 
C_{LP} \, (1-\beta)^{\gamma_{LP}} \, (1+a_{LP}(1-x_{\pp})^{\beta_{LP}}) \right\}
\nonumber \end{eqnarray} 
and similarly for gluons with the corresponding parameters $N_g$, $a_g$ and $b_g$.

Even though at the initial scale $Q^2_0=2.5\,$GeV$^2$ the parametrization implies some sort of flux factorization, beyond the initial scale, the evolution equations drive the fracture function as a whole making the usual discrimination between `fluxes' and `parton densities of the exchanged object'  somewhat ambiguous.

In order to ensure that the evolution code is working properly in the region of large $\beta$, we imposed a lower constraint on the $(1-\beta)$ exponents for both quark and gluon distributions,  $b_s, \, b_g > 0.1$.
Furthermore, in the expectation of very hard distributions we saturate this constraint and fix $a_s=0$, leaving only 8 free parameters. Doing this we obtain a global fit with $\chi_{total}^2/$d.o.f.=1.09 ($\chi^2=292.23$, data points $=274$), which we designate as Set A in Table~\ref{tab:exp}. 

Similar parametrizations, but with softer gluons, yield slightly higher values, for example in Set B, where $b_g$ is set to $0.7$ and $b_s$ remains 0.1,  finding $\chi_{total}^2/$d.o.f.=1.13.

As is shown by the solid lines in Figures. 1 and 2, the accuracy of the fit is remarkably good in the case of H1 diffractive data ($\chi_{H1D}^2/data=215.63/226$), and also good for the leading-proton data. The dashed line in figure 1 shows the contribution coming from the $x_{\pp}^{\alpha_{\pp}}$ term in Eq. (5), which could be interpreted as the `pomeronic component' of the fracture function. This contribution is clearly dominant for large $\beta$, but fails to do so in the low-$\beta$ and large-$x_{\pp}$ region.

\setlength{\unitlength}{0.6mm}
\begin{figure*}[hbt]
\begin{picture}(340,155)(0,0)
\put(13,-34){\mbox{\epsfxsize8.1cm\epsffile{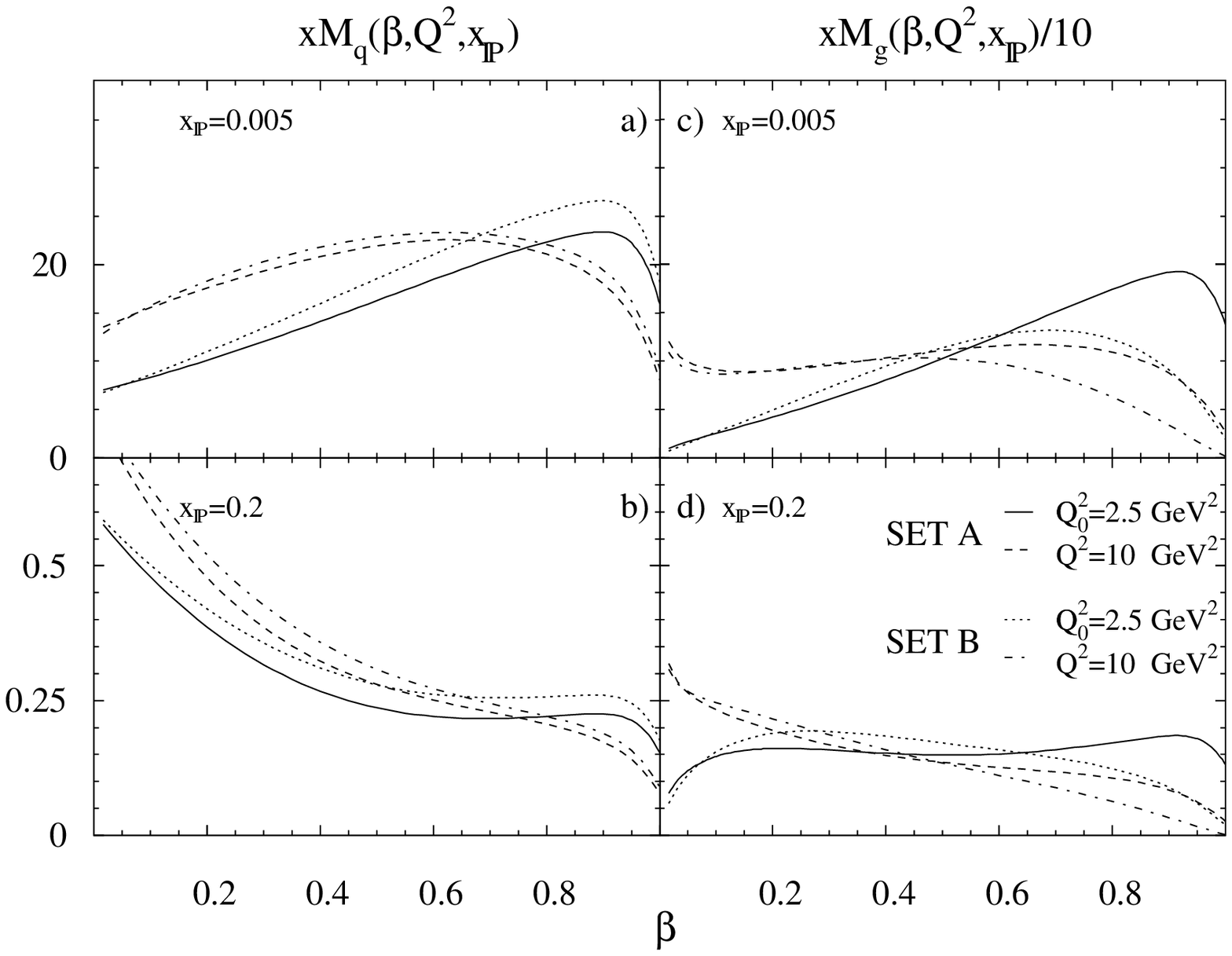}}}
\put(5,6){\mbox{\footnotesize Figure 3:  
Fracture function densities: a) and  b) light-quark }}
\put(5,0){\mbox{\footnotesize singlet, c) and d) gluons. }}
\put(175,-9){\mbox{\epsfxsize7.7cm\epsffile{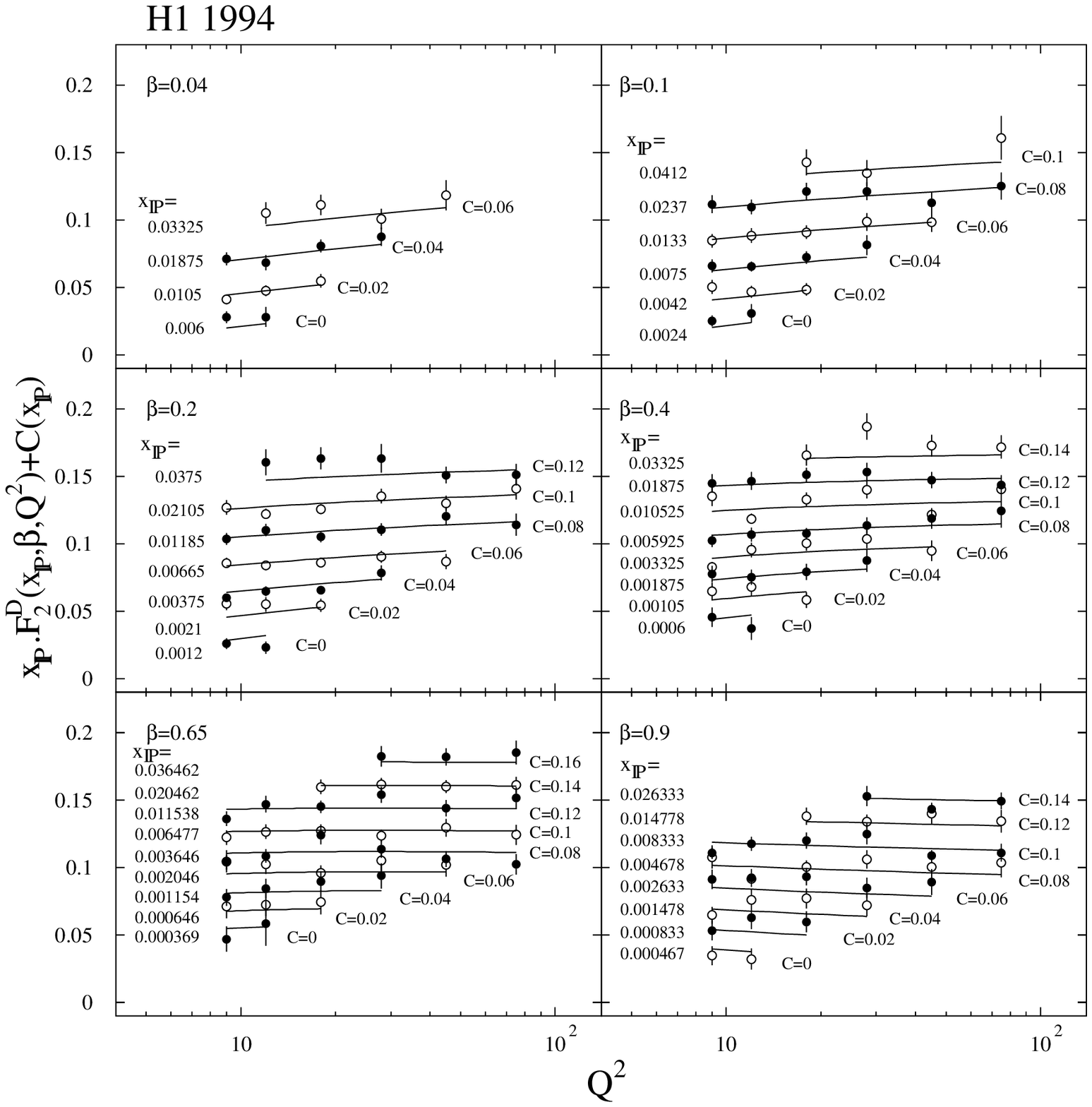}}}
\put(165,6){\mbox{\footnotesize Figure 4:  
Scale dependence of H1 diffractive data and the one}}
\put(165,0){\mbox{\footnotesize  obtained evolving the Set A parametrization.}}
\end{picture}
\end{figure*}

In the analysis of both H1 data sets \cite{H1D,H1LP}, which were obtained within different ranges of the variable $t$,
we have assumed the universal validity of the exponential behaviour in $t$ measured by the ZEUS collaboration \cite{ZD} and have rescaled the data to a common range, given by that of the diffractive data of H1 ($-M_p^2 x_{\pp}^2/(1-x_\pp) \equiv-t_{min}<-t<1$ GeV$^2$).

In Figure 3, we show the light-quark singlet and gluon  fracture densities at the initial scale $Q_0^2$  and at an intermediate value of $Q^2=10\,$GeV$^2$ for two characteristic values of $x_{\pp}$. The first one, $x_{\pp}=0.005$, which corresponds to well inside the diffractive regime, shows a rather hard behaviour, whereas the second one, $x_{\pp}=0.2$, which belongs to the 
leading-proton regime, is much softer. It is worth noticing that, at large $x_{\pp}$, the large-$\beta$ behaviour of the distributions is not well constrained because  leading-proton data are only available in the very small-$\beta$ range.

As it is also shown in Figure 3, gluons carry much more impulse than quarks, specially in the case of the diffractive regime and at low scales. The evolution damps down this gluon dominance, suggesting a large fraction of valence-like gluons
in the `pomeron', which is not so apparent for `reggeons'.  

The main conclusion that can be drawn about the gluon density from the   analysis is that   the
 distribution is important at large $\beta$, at variance with the one for inclusive structure functions, but  its behaviour cannot be precisely determined yet from the available data,  in particular, the exponent  of the $(1-\beta)$ factor in the parametrization.

\section{Scale dependence}

As it has been said, the fracture function approach leads to very definite predictions about the scale dependence of the  semi-inclusive cross sections. The rigorous factorization of the cross sections achieved within
this formalism allows a precise perturbative QCD analysis of the scale 
dependence of the data, as is usually done for ordinary structure functions.

In figures 4 and 5 we compare H1 diffractive and leading-proton data, at fixed values of $x_{\pp}$ and $\beta$, as a function of $Q^2$, with the evolved fracture function. The scale dependence induced by the evolution equations is perfectly consistent with the data.

It is also worth noticing that within the measured range, the scale dependence is dominated by the homogeneous evolution. The effects of the inhomogeneous term 
\setlength{\unitlength}{0.6mm}
\begin{figure}[hbt]
\begin{picture}(170,150)(0,0)
\put(15,-9){\mbox{\epsfxsize7.5cm\epsffile{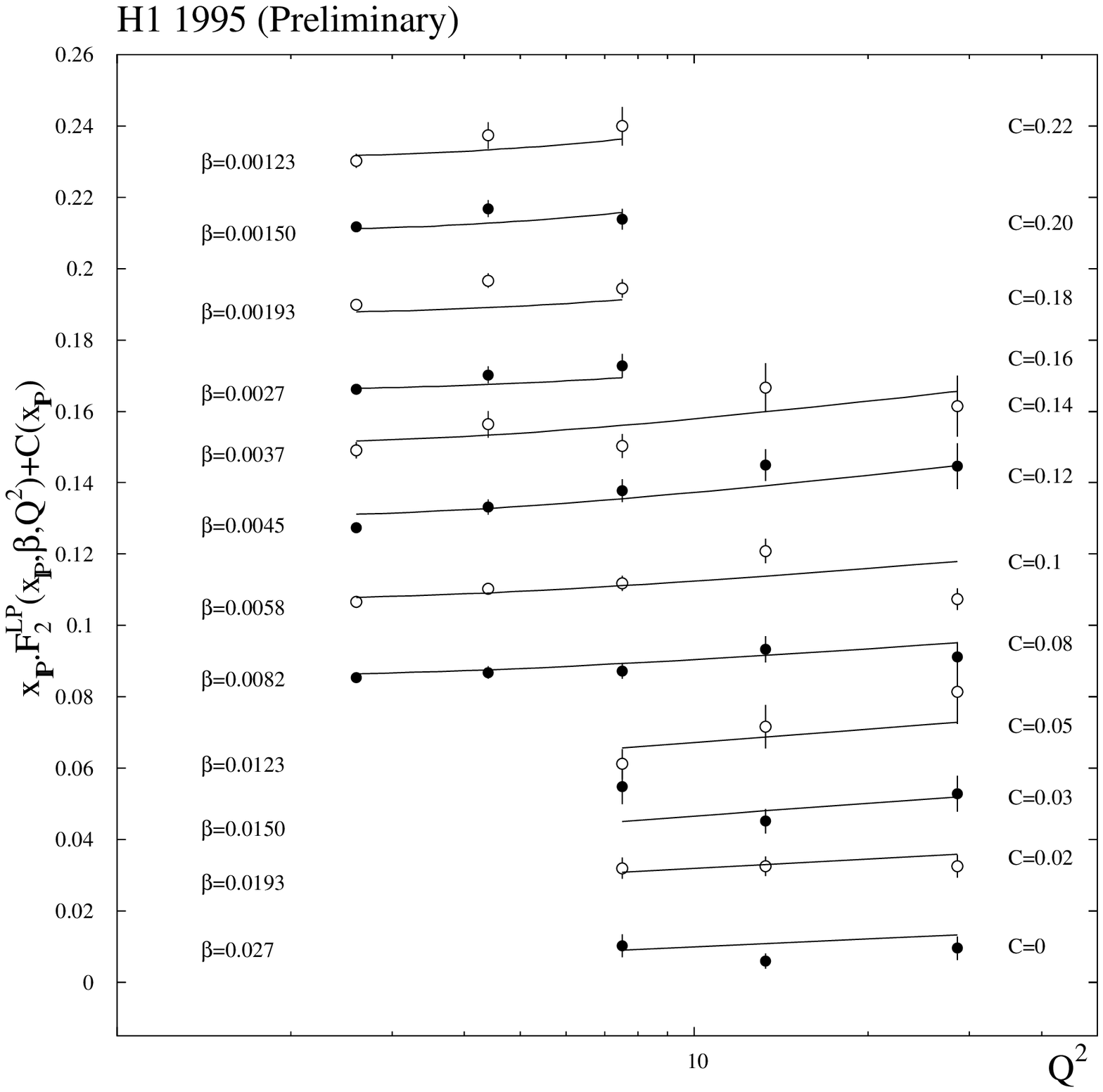}}}
\put(4,6){\mbox{\footnotesize Figure 5:  
Scale dependence of H1 leading proton data and the}}
\put(4,0){\mbox{\footnotesize one obtained evolving the Set A parametrization.}}
\end{picture}
\end{figure}

\noindent become important only for low values of $x_L$ where fragmentation functions are larger. These effects are, however, beyond the range of present data \cite{ph}.

\section{ZEUS measurements}

In order to check the distributions obtained in the previous section,in the following we compare results obtained using our best parametrization (Set A) with data sets presented by the ZEUS collaboration, which had not been included in the fit.

\setlength{\unitlength}{0.6mm}
\begin{figure}[hbt]
\begin{picture}(170,152)(0,0)
\put(15,0){\mbox{\epsfxsize7.cm\epsffile{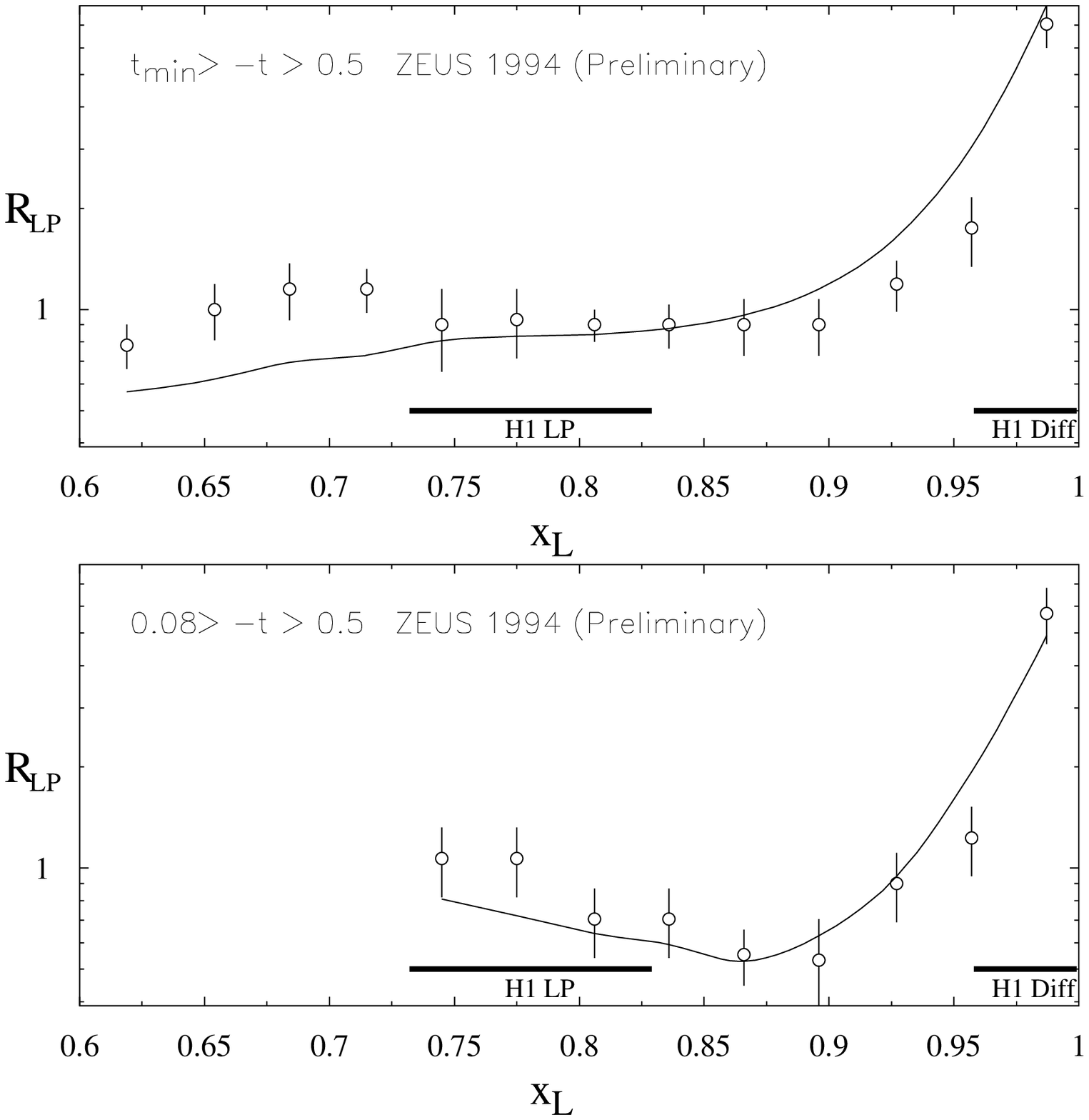}}}
\put(5,10){\mbox{\footnotesize Figure 6:  
The fraction of DIS events with a leading-proton for }}
\put(5,4){\mbox{\footnotesize different ranges of $t$ as measured by ZEUS}}
\end{picture}
\end{figure}

\setlength{\unitlength}{0.6mm}
\begin{figure}[hbt]
\begin{picture}(170,127)(0,0)
\put(15,-17){\mbox{\epsfxsize6.7cm\epsffile{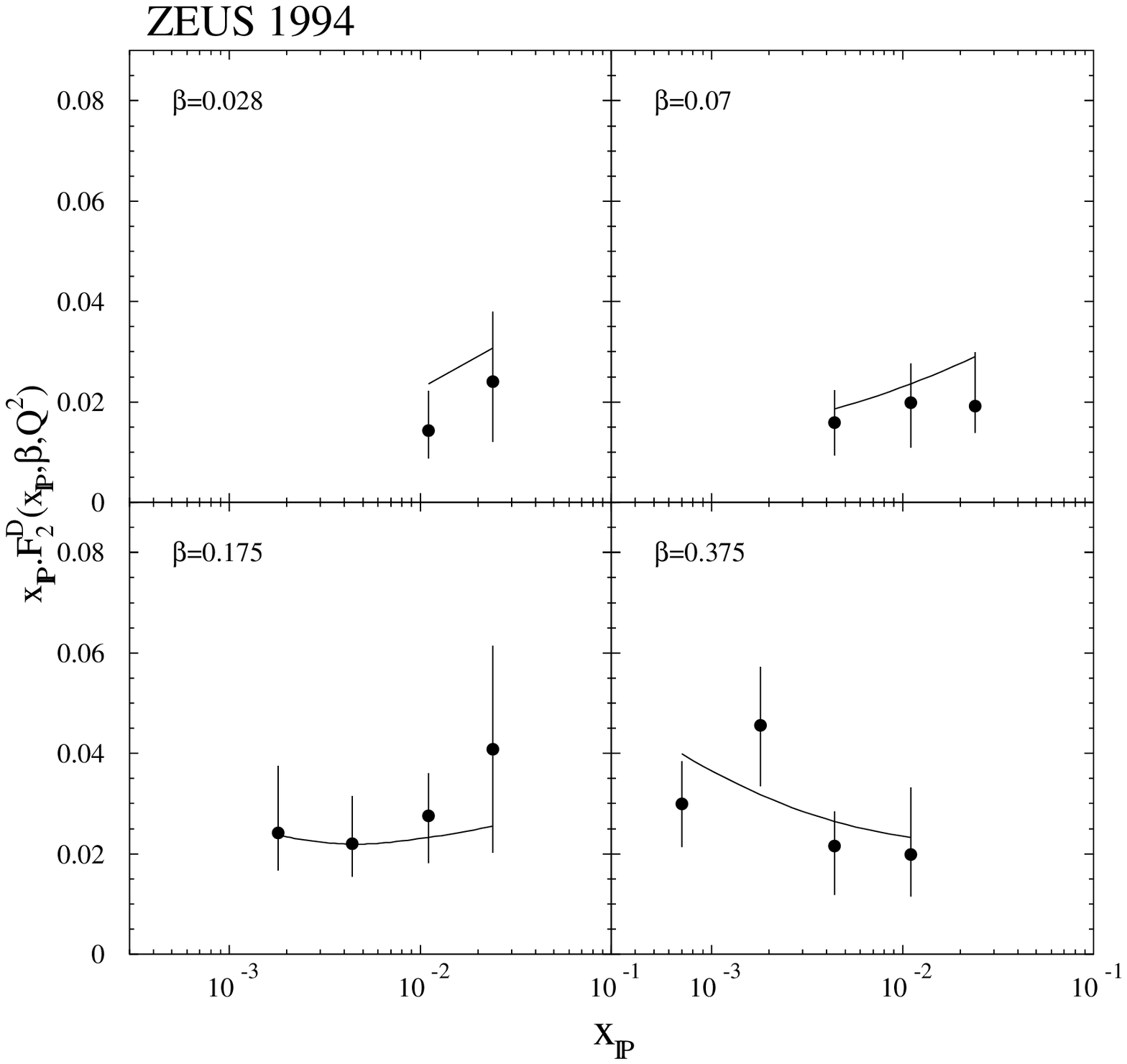}}}
\put(9,6){\mbox{\footnotesize  Figure 7:} {\footnotesize ZEUS diffractive data, against the expectation }}
\put(9,0){\mbox{\footnotesize  coming from the fracture function parametrization.}}
\end{picture}
\end{figure}

The ZEUS collaboration has measured the fraction of DIS events with a leading-proton in the final state\cite{ZLP}. After the adequate rescaling   of the parametrization for the fracture function at an average value of $Q^2=10$ GeV$^2$ we obtain a remarkable agreement with the data in the common $x_L$-range, as shown in Figures 6a and 6b. Notice also  the fact that the parametrization interpolates fairly well  the diffractive and leading-proton behaviour between H1 kinematical regions (thick lines) where the data used in the fit belong. 
In Figure 7 we compare ZEUS diffractive measurements \cite{ZD} with the outcome
of the parametrization after the appropriate evolution to the mean scale value of the data $Q^2=8 \,$GeV$^2$, and the already mentioned rescaling in $t$.

\section{Conclusions}

We have shown that an approach based on fracture functions motivates and allows a unified
description of both leading-proton and diffractive deep inelastic cross sections. A simple parametric form for this function gives a very accurate
description of the  data  available at present providing a smooth interpolation between the distinctive behaviours of the two regimes, also in accordance with ZEUS data. The analysis also hints at  some non-perturbative features,  such as a strong-gluon dominance in the `pomeronic' component with a characteristic valence-like behaviour. Finally, our results  verify that the scale dependence of the data agrees with the one
predicted by the fracture function formalism.

\end{document}